\documentclass[]{article}
\usepackage{jltp}
\usepackage{amssymb}
\usepackage{bookmath}
\usepackage{epsf}
\usepackage{psfig}
\usepackage{rotate}
\def\v#1{{\bf #1}}
\renewcommand{\vec}{\v}
\title{Magnetic Susceptibility of the Balian-Werthamer Phase of $^3$He in Aerogel}
\author{Priya Sharma and J.A. Sauls}
\address{Department of Physics \& Astronomy, Northwestern University, Evanston, IL 60208 \\
(July 5, 2001)}
\runninghead{P. Sharma and J. A. Sauls}{Susceptibility of $^3$He-B in Aerogel}    
\begin{document}
\begin{abstract}
\begin{center}
\begin{minipage}{0.9\textwidth}
\centerline{\bf Abstract}
The equilibrium superfluid phase of $^3$He impregnated into high-porosity 
silica aerogels appears to be a non-equal-spin-pairing 
state in zero field at all pressures, which is generally assumed to be the 
Balian-Werthamer (BW) phase modified by the depairing effects of the aerogel structure.
The nuclear magnetic susceptibility played a key role in identifying the B-phase
of pure $^3$He with the BW state. We report theoretical
calculations of the nuclear magnetic susceptibility for the BW model 
of superfluid $^3$He in aerogel within the framework of the Fermi-liquid 
theory of superfluid $^3$He. Scattering of quasiparticles by the aerogel,
in addition to Fermi-liquid exchange corrections, leads to substantial 
changes in the susceptibility of the BW phase. The increase in the magnetic
susceptibility of $^3$He-aerogel compared to pure $^3$He-B is related to 
the polarizability of the gapless excitations and the impurity-induced
local field. The limited data that is available is in rough
agreement with theoretical predictions. Future measurements could prove 
important for a more definitive identification of the ordered phase, as well as for 
refining the theoretical model for the effects of disorder and scattering 
on the properties of superfluid $^3$He.
\end{minipage}
\end{center}
\end{abstract}
\maketitle

\section{Introduction}

Early torsional oscillator and NMR experiments on $^3$He impregnated into 
silica aerogel provided the first evidence for a superfluid transition for $^3$He in 
a disordered medium. The transition temperature is suppressed in 98\% porous aerogel
to $T_c/T_c^{\rm bulk}\simeq 0.7$.\cite{por95,spr95}
Sizeable suppressions of the superfluid density and transverse NMR shift were
also reported. NMR measurements of the susceptibility indicated that an 
equal-spin-pairing (ESP) state was stable at pressures as low as 12 bar \cite{spr95},
suggesting that the equilibrium phase of $^3$He in aerogel at low pressures and
non-zero field was 
\underline{not} the Balian-Werthamer (BW) state that characterizes the bulk 
B-phase of superfluid $^3$He. However, with the addition of roughly four monolayers 
coverage of $^4$He on the aerogel strands, the magnetization was observed to decrease 
below the aerogel transition indicative of a non-ESP, B-like, pairing state.\cite{spr96}

More recent measurements have clarified the phase diagram of $^3$He in aerogel.
Torsional oscillator measurements
over the full pressure range ($0-34\,\mbox{bar}$) for temperatures as low as 0.1 mK 
provide convincing evidence of a zero-temperature, normal-superfluid phase transition 
as a function of pressure with a critical pressure of $P_c\simeq 6\,\mbox{bar}$.\cite{mat97}
Additional evidence for a B-like phase over the pressure range $P=1.5-29.3\,\mbox{bar}$ was
reported by Alles, et al. \cite{all99} based on the textural lineshape analysis of NMR
data on $^3$He in 98\% and 99\% porous aerogels. Barker et al. \cite{bar00} also reported 
the suppression of the magnetic susceptibility with two monolayers 
of $^4$He added to suppress the Curie component of the magnetization of solid $^3$He 
that plates to the aerogel strands. Their NMR measurements provide
evidence for an A-like phase very close to $T_c$ and a B-like (non-ESP) phase at slightly 
lower temperatures, with a first-order `AB' transition exhibiting substantial supercooling. 
The AB transition in $^3$He-aerogel was also observed at low pressures
($P=4.8\,\mbox{bar}$) in relatively high fields ($H_{\mbox{\tiny AB}}\simeq 0.2\,\mbox{T}$)
by Brussaard, et al. \cite{bru01}. Gervais, et al.\cite{ger01} have shown 
that sound attenuation measurements at frequencies of 
$\sim 15\,\mbox{MHz}$ provide a powerful tool to map the phase diagram of $^3$He-aerogel 
as a function of magnetic field, temperature and pressure. Their recent measurements of $T_c$ and
$T_{\mbox{\tiny AB}}$ as a function of field provide strong evidence for a stable B-like phase 
in the zero-field limit for pressures of $25$ and $33\,\mbox{bar}$.\cite{ger01,ger01b}
The AB transition is 
suppressed below the zero-field $T_c$ quadratically with field:
$(T_{\mbox{\tiny AB}}(H) - T_c)/T_c \simeq - 0.05\,H^2\,\mbox{kG}^{-2}$ at $P=25\,\mbox{bar}$.
These authors also find that the 
A-like phase exhibits strong supercooling even in low magnetic fields, which is 
the likely explanation for earlier observations of an ESP pairing state in the susceptibility
measurements by Sprague et al. \cite{spr95}.

\begin{figure}[h]
\begin{center}
\begin{minipage}{0.85\hsize}
\centerline{\psfig{file=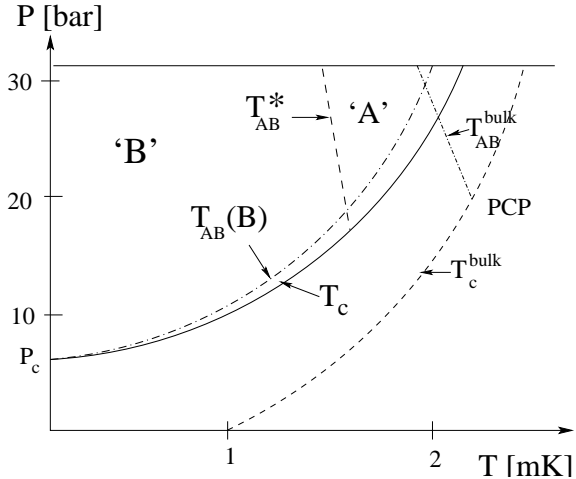,width=0.7\linewidth}}
\caption{\small Sketch of the phase diagram of superfluid $^3$He in 98\% 
         porous aerogel. The B-like phase extends to the highest
	 pressures. The quantum phase transition is shown at
	 $P_c\simeq 6\,\mbox{bar}$, and the field-induced AB
	 transition, $T_{\mbox{\tiny AB}}(H)$, is indicated 
	 by the dash-dotted line. Supercooled A-like $^3$He extends
	 down to $T^{*}_{\mbox{\tiny AB}}$. The bulk transition lines
	 are also indicated.}
\label{phase-diagram}
\end{minipage}
\end{center}
\end{figure}

The resulting phase diagram of superfluid $^3$He-aerogel (see Fig. \ref{phase-diagram})
differs significantly 
from that of bulk superfluid $^3$He. There is a zero-temperature, ``quantum'' phase 
transition at $P_c\simeq 6\,\mbox{bar}$ separating a disordered normal 
Fermi-liquid phase from a B-like superfluid phase. The A-like 
phase appears to be an equilibrium phase only in the presence of a magnetic field.
If there is a region of stable A-phase in zero-field then it is confined 
to a narrow temperature window below $T_c$ with 
$\vert T_{\mbox{\tiny AB}}(0)-T_c\vert \le 20\,\mu\mbox{K}$.\cite{ger01} 
The other significant feature of the AB transition in $^3$He-aerogel is 
the degree to which the A-like phase can be supercooled for pressures 
$P\gsim 20\,\mbox{bar}$.

The non-ESP `B-like phase' in aerogel is generally assumed to be the Balian-Werthamer
phase, modified by the aerogel structure. This identification is consistent with
theoretical expectations based on free energy calculations for the possible phases of 
superfluid $^3$He-aerogel within the homogeneous scattering model.\cite{thu98} 
Homogeneous, isotropic scattering leads to increased stabilization of the BW 
phase relative to the Anderson-Brinkman-Morel (ABM) phase, 
the planar phase and the polar phase. Nevertheless, the identification of the 
equilibrium phases of superfluid $^3$He in aerogel has not been rigorously 
established as it has been in bulk $^3$He.  Identification of the equilibrium 
phases of $^3$He-aerogel is complicated by uncertainties in the homogeneity 
and variations in the microstructure of aerogels of the same
density, the inhomogeneities induced into the order parameter by pairbreaking
effects of the aerogel structure, and the increased difficulty in carrying out
accurate theoretical calculations of the properties of $^3$He-aerogel.

\section{$^3$He-Aerogel Scattering Model}

In the homogeneous scattering model (HSM) the aerogel medium is represented by 
a random distribution of scattering centers.\cite{thu98} For $98\%$ porosity the 
typical diameter of the aerogel strands is $d\simeq 30\,\AA$ and the
mean distance between strands is $D\simeq 320\,\AA$, which
is very large compared to the Fermi wavelength, $k_f^{-1}\sim\AA$,
but comparable to or less than the bulk coherence length, 
$\xi_0=\hbar v_f/2\pi T_{c0}\simeq 200-800\,\AA$ over the pressure range
$P=0-34\,\mbox{bar}$. Thus, the aerogel
does not significantly modify the bulk properties of normal $^3$He,
beyond the formation of a couple of atomic layers of solid-like $^3$He
on the silica strands. The dominant effect of the aerogel structure is
to scatter $^3$He quasiparticles moving at the Fermi velocity.
Such scattering has dramatic effects on the formation and properties of the
superfluid phases. If the coherence length (pair size) is sufficiently 
long compared to the typical distance between scattering centers, then
the aerogel is well described by a homogeneous, isotropic scattering medium 
with a mean-free path determined by the aerogel geometry.

\begin{figure}[h]
\begin{center}
\begin{minipage}{0.85\hsize}
\centerline{\psfig{file=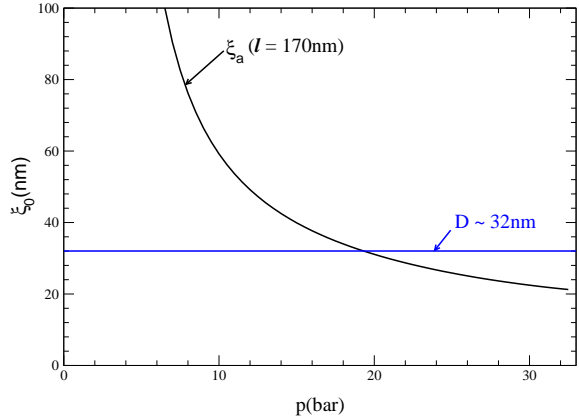,width=0.7\linewidth}}
\caption{\small The coherence length of superfluid $^3$He in 98\%
         porous aerogel as a function of pressure calculated within
	 the HSM for a mean-free path of $\ell = 1700\,\AA$. $D\simeq 320\,\AA$ is
         the typical distance between scattering centers.
	 }
\label{xi-pressure}
\end{minipage}
\end{center}
\end{figure}

Figure \ref{xi-pressure} shows the dependence of the Cooper
pair size, $\xi = \hbar v_f/2\pi T_c$, on pressure in superfluid $^3$He-aerogel 
calculated for aerogel with a mean free path of $\ell = 1700\,\AA$. 
The HSM may be expected to provide a reasonable first 
approximation to the properties of superfluid $^3$He-aerogel, particularly
at pressures below $15-20\,\mbox{bar}$. Indeed the HSM accounts semi-quantitatively
for the reduction of $T_c$, including the quantum critical pressure $P_c$, and
the pair-breaking suppression of the order parameter.\cite{thu98,rai98} 
However, the HSM is expected to become a poorer description of $^3$He-aerogel 
at higher porosities and higher pressures where the pair size becomes
comparable to, or smaller than, the mean distance between scattering centers. 
This breakdown of the HSM is evident in the quantitative discrepancies
in the pressure dependence of $T_c$ and $\rho_s/\rho$, particularly for higher 
porosity aerogels.\cite{law00}
More elaborate scattering models which take into account the similar scales 
for the pair correlation length and the aerogel correlation length, 
provide more accurate scaling behavior for the physical
properties of the superfluid phases with $\xi_0(P)$.\cite{thu98}

In the following we calculate the pairbreaking effects of quasiparticle scattering by
the aerogel, within the HSM,
on the nuclear magnetic susceptibility of the Balian-Werthamer phase
of superfluid $^3$He in aerogel. Comparison with existing data for the magnetization is also
presented. Extensions of this theory to include aerogel correlation effects and inhomogeneities
in the aerogel medium can also be carried out if future measurements warrant.

\section{Magnetization of Superfluid $^3$He-B-aerogel}

The nuclear magnetization, $M$, of normal liquid $^3$He at temperatures, $k_B T\ll E_f$, 
and fields, $\gamma\hbar H \ll E_f$, is given in terms of the (single-spin) 
quasiparticle density of states at the Fermi level, $N_f$, the nuclear gyromagnetic 
ratio for $^3$He, $\gamma$, and the exchange enhancement of the local field given 
in terms of the Landau interaction parameter, $F_0^a$,
\be
\chi_N = M/H  = \frac{2N_f\mu^2}{1+F_0^a}
\,,
\ee
where $\mu=\gamma\hbar/2$ is the nuclear magnetic moment of the $^3$He nucleus;
$\chi_N$ is the nuclear spin susceptibility of the normal Fermi liquid.

The effect of the aerogel on the magnetization of the normal liquid phase of $^3$He
is expected to negligible. However, the aerogel structure is known to be covered
with one or two layers of localized $^3$He atoms. These surface solid layers
contribute a Curie-like susceptibility that obscures the Fermi-liquid contribution
at low temperatures.\cite{spr95} The surface contribution can be suppressed 
by the addition of a couple of monolayers of $^4$He which preferentially plates 
the aerogel structure. The net effect is two-fold: (1) the surface Curie susceptibility
is suppressed {\sl and} (2) spin-spin scattering between $^3$He quasiparticles and
the surface spins is suppressed. The cross-section of the aerogel may also be 
modified by $^4$He preplating, but we expect this effect to be relatively small.
Measurements of the B-like suppression of the
susceptibility in $^3$He-aerogel have so far been reported only for two or more
monolayers of $^4$He added to displace the solid layer of $^3$He.\cite{spr95,bar00}
In the following we assume $^4$He coats the aerogel surface and consider only
{\sl nonmagnetic} scattering of $^3$He quasiparticles off the aerogel structure.

The nuclear spin susceptibility of pure superfluid $^3$He-B agrees quantitatively to leading
order in $T_c/E_f$ with the result of Serene and Rainer\cite{ser83} for the 
susceptibility of the Balian-Werthamer state,
\be
\label{chi-bulk}
\chi_B/\chi_N = \frac{(1+F_0^a)\left[\twothirds + Y(\onethird + \onefifth F^a_2)\right]}
                     {1+F_0^a(\twothirds+\onethird Y)
		     +\onefifth F^a_2(\onethird+\twothirds Y)
                     +\onefifth F^a_2 F^a_0 Y}
\,,
\ee
where $Y(T)$ is the well-known Yosida function,
\be
\label{Yosida}
Y(T) = 1 - \pi T \sum_{\varepsilon_n}\,
\frac{\Delta^2}{\left[\varepsilon_n^2+\Delta^2\right]^{3/2}}
\,,
\ee
and $\Delta(T)$ is the B-phase gap amplitude.
Indeed this result, and its generalization to include nonlinear field corrections arising from the
pair-breaking effect of the nuclear Zeeman energy\cite{sch82a,fis86}, has been used to obtain
the $\ell=2$ Landau parameter, $F_2^a$, from measurements of the susceptibility and gap distortion
of the B-phase collective modes.\cite{hoy81}

Below the superfluid transition in aerogel the magnetization is given by
\be
\label{magnetization}
\vM=\chi_N\left(\vH+\frac{\vm}{2\mu}\right)
\,,
\ee
where $\vm$ is related to a Fermi surface average of the change in spin polarization 
of quasiparticles at the point $\hat\vp$ on the Fermi surface,
\be\label{depolarization}
\vm=2T\sum_{\varepsilon_n}\,\int\frac{d\Omega_{\hat\vp}}{4\pi}\,\vec{g}(\hat\vp;\varepsilon_n)
\,.
\ee

The change in spin polarization leads to a reduction in the magnetization;
$\vm/2\mu=-D(T)\vH$, where $D(T)$ is the net depolarization below $T_c$.
The change in the spin-polarization is determined by competing effects 
of pairing correlations of quasiparticles with $S_z=0$, and pairbreaking 
induced by scattering off the aerogel structure. The effects of depairing
of the $S_z=0$ Cooper pairs by scattering from the aerogel are expected to be
significant and to lead to sizeable changes in the magnetization.
An obvious effect of scattering by the aerogel 
on the magnetization compared to that of bulk $^3$He-B is the suppression of  
$T_{c}$ relative to $T_{c0}$. However, the magnitude of the susceptibility,
particularly at low temperatures, is sensitive to the density of quasiparticle
states below the gap, $\varepsilon < \Delta$, produced by scattering off the aerogel.

In the HSM for {\sl isotropic scattering} in the Born
and unitarity scattering limits the generic form of Eq. (\ref{chi-bulk}) for the B-phase
susceptibility is preserved with the replacement of the gap and Yosida functions
by impurity-renormalized gap and response functions.
The results can be summarized by Eq. (\ref{chi-bulk}) with the replacement of 
$Y(T)\rightarrow \tilde{Y}(T)$ in the Born and 
unitarity scattering limits,
\ber
\label{Y-Born}
\tilde{Y} &=& 1 - \pi T\sum_{n}\,
                   \frac{\Delta^2}{\left[\tilde{\varepsilon}_n^2+\Delta^2\right]^{3/2}}
         \Bigg\{\frac{1}{1- \tinyonethird
		   \frac{\Gamma_N\Delta^2}{\left[\tilde{\varepsilon}_n^2+\Delta^2\right]^{3/2}}}
	 \Bigg\}
\quad(\mbox{Born})
\,,
\\
\label{Y-unitarity}
\tilde{Y} &=& 1 - \pi T\sum_{n}\,
                     \frac{\Delta^2}{\left[\tilde{\varepsilon}_n^2+\Delta^2\right]^{3/2}}
         \Bigg\{\frac{1}{1+ \tinyonethird
	                   \left(\frac{1}{\tilde{\varepsilon}_n}\right)^2
			   \frac{\Gamma_N\Delta^2}{\sqrt{\tilde{\varepsilon}_n^2+\Delta^2}}}
	\Bigg\}
\quad(\mbox{unitarity})
\,,
\eer
where $\Gamma_N$ is related to the mean-free path ($\ell$), or scattering rate ($1/\tau$), for 
normal-state $^3$He quasiparticles by
\be
\Gamma_N=\frac{\hbar}{2\tau}=\frac{\hbar v_f}{2 \ell}
\,,
\ee
and the gap and renormalized Matsubara frequencies are 
defined in Eqs. (\ref{gap-aerogel}) and (\ref{epsilon-tilde}).
The terms that depend explicitly on the scattering rate arise from impurity vertex corrections
to the local field. The impurity field, which has opposite signs in the Born and Unitary limits,
provides a small, but non-negligible, correction to the depolarization at low temperatures.
These results are derived in the next section, and discussed 
quantitatively in Sec. (\ref{Results}). 

\section{Fermi-Liquid Theory for Superfluid $^3$He-aerogel}

The properties of superfluid $^3$He in aerogel can be
calculated within the framework of the Fermi-liquid theory of superfluid $^3$He.\cite{ser83}
The effects of scattering by the aerogel are included within the HSM.\cite{thu98} 
The central equation of the Fermi-liquid theory of superfluid $^3$He is the 
quasiclassical transport equation,
\be\label{QCtransport}
\left[i\varepsilon_n\hat{\tau}_3 - 
       \hat{\Delta} -
       \hat{\Sig}_{\mbox{\tiny imp}} -
       \hat{\Sig}_{\mbox{\tiny FL}} -
       \hat{v}_{\mbox{\tiny ext}} 
       \,,\,
       \hat{\g}(\hat{\vp},\vR;\varepsilon_n)
\right] +i\vv_f\cdot\grad\hat{\g} = 0
\,,
\ee
where $\varepsilon_n$ are the Fermion Matsubara frequencies, $\vv_f=v_f\hat{\vp}$ is the 
Fermi velocity for excitations near the point $\hat{\vp}$ on the Fermi surface
and $\hat{\g}$ is the quasiclassical Green's function, which is the $4\times 4$ Nambu matrix
in particle-hole/spin space. For a full discussion of the quasiclassical theory of superfluid
$^3$He, including the matrix notation used here, see the review by Serene and Rainer.\cite{ser83} 
The matrix structure of the propagator in particle-hole space is given by
\be
\label{QCpropagator}
\hat{\g} = 
\left(
\begin{array}{cc}
g + \vec{g}\cdot\vsigma & i\sigma_2 f + i\vsigma\sigma_2\cdot\vec{f} \\
i\sigma_2\underline{f}+\underline{\vec{f}}\cdot i\sigma_2\vsigma & 
\underline{g}+i\sigma_2\,\underline{\vec{g}}\cdot\vsigma\,i\sigma_2
\end{array} 
\right)
\ee
where $\sigma_i\,\,(i=1,2,3)$ are the Pauli spin matrices, $g$ and $\vg$ are the
spin scalar and vector components of the diagonal quasiclassical propagator, and
$f$ and $\vf$ are the off-diagonal spin-singlet and triplet pair amplitudes.
The diagonal Matsubara Green's function determines the density of states,
magnetization, local exchange field etc. 
Fundamental symmetry relations connect the particle and hole components of
the propagators and self-energies,
\ber
\label{gsymmetry}
\underline{g}(\hat\vp,\varepsilon_n)=g(-\hat\vp,-\varepsilon_n)      \,&,&\quad
\underline{\vg}(\hat\vp,\varepsilon_n)=\vg(-\hat\vp,-\varepsilon_n)  \,,\\ 
\label{fsymmetry}
\underline{f}(\hat\vp,\varepsilon_n)=f(-\hat\vp,\varepsilon_n)^*     \,&,&\quad
\underline{\vf}(\hat\vp,\varepsilon_n)=-\vf(-\hat\vp,\varepsilon_n)^*\,.
\eer

The gradient term in Eq. (\ref{QCtransport}) determines the range for spatial 
variations of the pair amplitude and quasiparticle excitations in the superfluid 
phases. In the following we consider homogeneous equilibrium states, in which
case we drop the gradient term in the transport equation.
The matrix terms in Eq. (\ref{QCtransport}) represent different physical contributions to the 
self-energy and coupling to external fields. The self-energy 
terms can be classified in terms of the expansion parameters of Fermi liquid theory,
e.g. $k_B T_c/E_f$, $1/k_f\xi_0$, $\hbar/\tau E_f$, $\mu H/E_f$, etc.\cite{ser83}
The leading order contributions represent the off-diagonal pairing self energy, 
$\hat{\Delta}$, the self-energy resulting from multiple scattering of quasiparticles 
by the aerogel, $\hat{\Sig}_{\mbox{\tiny imp}}$, the Landau molecular field energies, 
$\hat{\Sig}_{\mbox{\tiny FL}}$, and the Zeeman coupling of the nuclear moments 
to an external magnetic field, $\hat{v}_{\mbox{\tiny ext}}$.

The off-diagonal self energy defines the order parameter; in matrix form,
\be\label{Delta-Nambu}
\hat{\Delta}(\hat{\vp}) \equiv
\left(\matrix{ 0 & \Delta \cr \underline{\Delta} & 0 }\right) =
\left(\matrix{ 0 & i\vsigma\sigma_y\cdot\vDelta(\hat{\vp})
               \cr 
	       i\sigma_y\vsigma\cdot\underline{\vDelta}(\hat{\vp}) & 0 }\right)
\,,
\ee
where we consider pure p-wave, spin-triplet pairing, which is sufficient for 
investigating the linear response to the nuclear Zeeman field.\cite{fis86} 
The order parameter $\vDelta$ satisfies the weak-coupling gap equation,
\be
\label{gapequation}
\vec{\Delta}(\hat\vp) = T\,\sum_{\varepsilon_n}^{|\varepsilon_n|\le\epsilon_c}\,
                             \int \frac{d\Omega_{\hat\vp'}}{4\pi}
			     V^t(\hat\vp\cdot\hat\vp')\,
			     \vec{f}(\hat\vp';\varepsilon_n)
\,,
\ee
where the pairing interaction in the spin-triplet, p-wave channel is given by,
$V^t = 3 V_1 \hat{\vp}\cdot\hat{\vp}'$. The interaction, $V_1$, and cutoff, $\epsilon_c$,
that enter Eq. (\ref{gapequation}) can be eliminated in favor of the measured bulk transition
temperature using the linearized equilibrium gap equation for 
pure $^3$He in zero field (see below).

In the HSM the scattering of $^3$He quasiparticles off the aerogel structure is modeled by
a random distribution of scattering centers (``impurities''). The 
impurity self-energy to leading order in $\hbar/\tau E_f$ is determined by the t-matrix
for multiple scattering by a single impurity and the mean density of impurities,
\be
\label{impurity-selfenergy}
\hat\Sig^{\mbox{\tiny imp}}(\hat\vp;\varepsilon_n)=
n_{\mbox{\tiny imp}}\,\hat{t}(\hat\vp,\hat\vp;\varepsilon_n) 
\,,
\ee
where $n_{\mbox{\tiny imp}}$ is the impurity density
and $\hat{t}$ is obtained from the self-consistent solution to
the t-matrix equation,
\be
\label{t-matrix}
\hat{t}(\hat\vp,\hat\vp';\varepsilon_n)=\hat{u}(\hat\vp, \hat\vp')+ 
N_f\,\int\frac{d\Omega_{\hat\vp''}}{4\pi}\,
                       \hat{u}(\hat\vp,\hat\vp'')\,
                       \hat{\g}(\vp'';\varepsilon_n)\,
		       \hat{t}(\hat\vp'',\hat\vp';\varepsilon_n)
\,,
\ee
the quasiclassical transport equation for $\hat{\g}$ and the constitutive equations
for the self energies.
The general structure of the impurity self-energy matrix is given by
\be\label{Sigma_imp}
\hat{\Sig}_{\mbox{\tiny imp}}(\hat{\vp};\varepsilon_n) = 
     \left(\matrix{
           \Sig_{\mbox{\tiny imp}}
	 + \vh_{\mbox{\tiny imp}}\cdot\vsigma &
	   \Delta_{\mbox{\tiny imp}}\,i\sigma_y
	 + \vDelta_{\mbox{\tiny imp}}\cdot i\vsigma\sigma_y         \cr 
	   \underline{\Delta}_{\mbox{\tiny imp}}\,i\sigma_y
	 + \underline{\vDelta}_{\mbox{\tiny imp}}\cdot i\sigma_y\vsigma    & 
	   \underline{\Sig}_{\mbox{\tiny imp}}
         + \underline{\vh}_{\mbox{\tiny imp}}\cdot\vsigma^{\mbox{\tiny tr}}
    }\right)
\,,
\ee
where the same symmetry relations in Eqs. (\ref{gsymmetry}-\ref{fsymmetry})
relate the particle-hole components of the self-energy.

The matrix $\hat u(\hat\vp,\hat\vp')$ represents the effective impurity
potential for scattering of quasiparticles from $\vp\rightarrow\vp'$ on the Fermi surface.
In the simplest formulation of the HSM we neglect spin-flip scattering by the aerogel,
assume that the aerogel is isotropic on the 
coherence length scale and retain only the isotropic scattering potential.
The neglect of spin-flip scattering is justified for $^3$He-aerogel with a small 
concentration of $^4$He added to displace the $^3$He solid layers. The resulting 
scattering potential is then modeled by a single s-wave matrix element,
$\hat u(\hat\vp,\hat\vp')=u_0\hat{1}$. The assumption of isotropic scattering 
is more questionable. However, anisotropic 
scattering and orientational correlations of the anisotropic scattering centers
representing the silica strands can be incorporated into the HSM if needed.

The Landau molecular-field self-energy
\be
\hat{\Sig}_{\mbox{\tiny FL}}(\hat{\vp}) \equiv 
\left(\matrix{\Sig_{\mbox{\tiny FL}}+\vh(\hat{\vp})\cdot\vsigma & 0 \cr 
		0 & \underline{\Sig}_{\mbox{\tiny FL}}
		   +\underline{\vh}(\hat{\vp})\cdot\vsigma^{\mbox{\tiny tr}}
	     }
\right)
\,,
\ee
is a functional of the diagonal components of the quasiclassical propagator.
The scalar and vector components represent internal fields generated by
the quasiparticle interactions through the disturbance of the quasiparticle
spectrum in response to pairing correlations or external fields. The scalar
molecular field is given by
\be\label{FL-scalar}
\Sig^{\mbox{\tiny MF}}(\hat\vp)=T\,\sum_{\varepsilon_n}\,
                                   \int \frac{d\Omega_{\hat\vp'}}{4\pi}\,
				   A^s(\hat\vp\cdot\hat\vp')\,g(\vp';\varepsilon_n)
\,,
\ee
where $A^s(\hat\vp,\hat\vp')$ is the spin-independent quasiparticle-quasiparticle 
scattering amplitude. Similarly, the vector component of the molecular field, 
\be\label{FL-vector}
\vh(\hat\vp) = T\,\sum_{\varepsilon_n}\,
                  \int\frac{d\Omega_{\hat\vp'}}{4\pi}\,
                  A^a(\hat\vp\cdot\hat\vp')\,\vec{g}(\hat\vp';\varepsilon_n)
\,,
\ee
represents the internal exchange field acting on quasiparticles at the position $\hat\vp$
on the Fermi surface, and is determined by the quasiparticle-quasiparticle 
exchange scattering amplitude, $A^a(\hat\vp,\hat\vp')$. These forward scattering 
amplitudes are related to the Fermi-liquid parameters, $F^{(s,a)}_{\ell}$, through the 
Legendre expansion,\cite{lan57}
\be
A^{(s,a)}=\sum_{\ell}\frac{F_{\ell}^{(s,a)}}{1+F_{\ell}^{(s,a)}/2\ell + 1}\,
          P_{\ell}(\hat\vp\cdot\hat\vp')
\,.
\ee
In calculating the linear response to a uniform magnetic field only the
exchange interactions, $F_0^a$ and $F_2^a$, contribute for isotropic impurity
scattering in the Born and unitarity scattering limits.

Finally, we have the Zeeman coupling to an external magnetic field, 
\be
\hat{v}_{\mbox{\tiny ext}}=\vh_{\mbox{\tiny ext}}\cdot\hat{\vSigma}
\,,\quad\mbox{with}\quad
\vec{h}_{\mbox{\tiny ext}}=-Z_0^a\,\mu\vH
\,,
\ee
$\hat{\vSigma}=
\tinyonehalf(1+\hat{\tau}_3)\vsigma+\tinyonehalf(1-\hat{\tau}_3)\vsigma^{\mbox{\tiny tr}}$,
and $\mu=\gamma\hbar/2$ is the nuclear magnetic moment of the $^3$He nucleus.
The factor $Z_0^a = 1/(1+F_0^a)$ is the high-energy renormalization of the Zeeman
coupling of the quasiparticle moment to the external field. Spin-rotation invariance
in normal liquid $^3$He fixes the renormalization factor in terms of the quasiparticle exchange
interaction, $F_0^a$.\cite{ser83}

The quasiclassical transport equation and constitutive equations for the self-energies 
are supplemented  by Eilenberger's normalization condition for the matrix 
propagator,\cite{eil68}
\be
\hat{\g}^2 = -\pi^2 \hat{1}
\,.
\ee
In addition to the overall normalization of the propagator, the normalization condition 
is useful in simplifying the linearized transport equation for the response to an 
external field.

\subsection{Homogeneous equilibrium in zero field}

For the Balian-Werthamer phase the order parameter is given by
\be
\hat\Delta(\hat\vp) = \Delta\,\underline{R}\cdot\hat\vp
\,,
\ee
where $\underline{R}$ is the relative spin-orbit rotation matrix 
that defines the B-phase.\cite{leg75} The BW state is isotropic under joint
spin and orbital rotations; as a result the magnitude of the order parameter,
$|\vDelta(\hat\vp)|=\Delta(T)$, is also isotropic. The magnetic susceptibility 
is determined by exchange interactions and nonmagnetic scattering; 
we can safely neglect nuclear dipolar interactions, and for the purpose of calculating 
the susceptibility we can set $\underline{R}=1$.

The homogeneous equilibrium solution for the propagator and self-energies 
of the bulk BW phase is straight-forward to generalize for isotropic scattering, 
and has been investigated by Buchholz and Zwicknagl for the BW model for 
a superconductor.\cite{buc81} The molecular field corrections vanish in zero field,
so we are left with the mean-field order parameter and impurity self energy in the
homogeneous transport equation,
\be
\label{QCtransport_zerofield}
\left[i\varepsilon_n\hat\tau_3-\hat\Sig_{\mbox{\tiny imp,0}}-\hat\Delta_0\,,\,\hat{\g}_0\right]=0
\,, 
\ee
where $\hat{\g}_0$ is the equilibrium propagator in zero field. The solution 
for $\hat{\g}_0$ for an isotropic BW phase with isotropic scattering has the same matrix form
as that for the pure BW phase,
\be\label{g0_solution}
\hat{\g}_0=-\pi\,\frac{i\tilde\varepsilon_n\hat\tau_3-\hat{\tilde{\Delta}}_0(\hat\vp;\varepsilon_n)}
                      {\sqrt{\tilde\varepsilon_n^2+|\tilde{\vDelta}_0|^2}}
\,,
\ee
where the renormalized Matsubara frequencies and order parameter are related to
zero-field impurity self energy by (we drop the subscript `0' unless it is explicitly needed)
\be
i\tilde{\varepsilon}_n = i\varepsilon_n - \Sig_3(\varepsilon_n)
\,,
\ee
\be
\tilde{\vDelta}(\hat\vp;\varepsilon_n)=\vDelta(\hat\vp)+\vDelta_{\mbox{\tiny imp}}(\varepsilon_n)
\,,
\ee
with the zero-field impurity self energy given by,
\be
\hat{\Sig}_{\mbox{\tiny imp}}= \Sig_1(\varepsilon_n)\,\hat{1} +  
                                 \Sig_3(\varepsilon_n)\,\hat{\tau}_3 +  
                                 \hat{\Delta}_{\mbox{\tiny imp}}(\varepsilon_n)
\,.
\ee
However, the off-diagonal impurity self energy, $\vDelta_{\mbox{\tiny imp}}(\varepsilon_n)$,
vanishes for pure s-wave scattering because the 
the driving term in the equation for $\tilde{\vDelta}(\hat\vp;\varepsilon_n)$ is the mean
field order parameter $\vDelta(\hat\vp)$, which vanishes when averaged over
the Fermi surface,\cite{cho89b}
\be
\langle\vDelta(\hat\vp)\rangle = 
\int\frac{d\Omega_{\hat\vp}}{4\pi} \hat{\Delta}(\hat\vp) = 0
\,.
\ee
Thus, $\tilde{\vDelta}=\vDelta(\hat\vp)$, and  the t-matrix is isotropic and reduces
to diagonal form,
\ber
\label{t0}
\hat{t} &=& t_1\hat{1}+t_3\hat{\tau}_3 \,,
\\
\label{t01}
t_1&=&\frac{\sqrt{\bar{\sigma}(1-\bar{\sigma})}}{\pi N_f}\,
\frac{\tilde\varepsilon_n^2+\Delta^2}{\tilde\varepsilon_n^2+\Delta^2(1-\bar{\sigma})}\,,
\\
\label{t03}
t_3&=&-\frac{\bar{\sigma}}{\pi N_f}\,
\frac{i\tilde\varepsilon_n\,\sqrt{\tilde\varepsilon_n^2+\Delta^2}}
     {\tilde\varepsilon_n^2+\Delta^2(1-\bar{\sigma})}
\,,
\eer
where $\bar\sigma$ is the normalized scattering cross-section and is related to
the scattering potential, $u_0$, and scattering phase shift, $\delta_0$, by
\be
\bar{\sigma} = \sin^2\delta_0 = \frac{(\pi N_f u_0)^2}{1+(\pi N_f u_0)^2}
\,.
\ee
The unitarity limit corresponds to strong scattering, $u_0\rightarrow\infty$, 
in which case $\bar\sigma\rightarrow 1$ ($\delta_0\rightarrow\pi/2$). Born scattering
is the limit of small cross section, $\bar\sigma\ll 1$, ($\delta\ll 1$).
The other key parameter needed 
to relate the HSM parameters to the properties characterizing the aerogel is the mean
free path, $\ell$. In the HSM the mean-free path is determined by the 
Fermi velocity and the mean time for a quasiparticle to scatter off an impurity,
$\ell=v_f\tau$. The scattering time is determined by the 
imaginary part of the normal-state impurity self energy,
\be\label{scattering rate}
\Gamma_N=\frac{\hbar}{2\tau} = \frac{n_{\mbox{\tiny imp}}}{\pi N_f}\,\bar{\sigma}
\,.
\ee
Thus, the isotropic HSM is defined by two parameters which we take to be the mean
free path, $\ell$, and identify with the geometric mean free path for 
a specific aerogel; and the dimensionless cross-section of the scattering centers,
$\bar\sigma$, which can vary from $\bar\sigma=0$ (Born limit) to $\bar\sigma=1$ 
(unitarity scattering limit).
Since the geometric cross-section of a typical aerogel strand
is large compared to the Fermi wavelength the unitarity limit is more appropriate
for modeling the effects of scattering by aerogel. 

Finally, the renormalized Matsubara frequencies are given by the solution to
the equation,
\be\label{epsilon-tilde}
\tilde\varepsilon_n = \varepsilon_n + \Gamma_N\,\,
{\frac {\tilde\varepsilon_n\,\sqrt{\tilde\varepsilon_n^2 + \Delta^2}}
{\tilde\varepsilon_n^2 + \Delta^2(1-\bar\sigma)}}
\,.
\ee
The equilibrium equations in zero field are closed by the self-consistency equation 
for the order parameter,
\be
\vDelta(\hat\vp) = \pi T\sum_{\varepsilon_n}^{|\varepsilon_n|\le\epsilon_c}\,
                   \int\frac{d\Omega_{\hat\vp'}}{4\pi}\,V^t(\hat\vp\cdot\hat\vp')\,
		   \frac{\vDelta(\hat\vp')}{\sqrt{\tilde{\varepsilon}_n^2 + \Delta^2}}
\,,
\ee
which reduces to the BCS gap equation,
\be
\Delta = V_1\,\pi T\sum_{\varepsilon_n}^{|\varepsilon_n|\le\epsilon_c}\,
		   \frac{\Delta}{\sqrt{\tilde\varepsilon_n^2 + \Delta^2}}
\,.
\ee
The linearized gap equation for pure $^3$He-B ($\tilde{\varepsilon}_n\rightarrow\varepsilon_n$)
determines the bulk transition temperature,
\be\label{Tc0}
1/V_1 = \pi T_{c0}\sum_{\varepsilon_n}^{|\varepsilon_n|\le\epsilon_c}\,\frac{1}{|\varepsilon_n|}
      = K(T_{c0}) \simeq \ln(1.13\epsilon_c/T_{c0})
\,.
\ee
Using Eq. (\ref{Tc0}) we can remove the pairing interaction and cutoff in favor of the
bulk transition temperature, $T_{c0}$. Thus, the gap equation for $^3$He-B-aerogel becomes,
\be\label{gap-aerogel}
\ln(T/T_{c0})=
\pi T\sum_{\varepsilon_n}\left(\frac{1}{\sqrt{\tilde{\varepsilon}_n^2+\Delta^2}}-\frac{1}{|\varepsilon_n|}\right)
\,.
\ee
The linearized gap equation reduces to the Abrikosov-Gorkov formula\cite{abr59a} for 
the suppression of $T_c$ by scattering in the HSM for the aerogel,\cite{thu98}
\be
\ln(T_{c}/T_{c0})=2\pi T_c \sum_{n\ge 0}\left(\frac{1}{\varepsilon_n+\Gamma_N}-
                                              \frac{1}{\varepsilon_n}\right)
\,.
\ee

\subsubsection*{Density of States}

In superconductors with an order parameter that breaks orbital rotation symmetry, 
such as d-wave pairing in the cuprates or f-wave pairing in UPt$_3$, scattering from an impurity, 
particularly in the strong scattering limit, or from a surface, leads to the formation 
of quasiparticle states at the Fermi level that are bound to the impurity, or surface,
within a distance of order the coherence length.\cite{hu94,bal95} In the case of a random 
distribution of scattering centers the impurity bound states at the Fermi level broaden into 
an {\sl impurity band} near the Fermi level.\cite{buc81} The same is true for the 
p-wave phases of $^3$He in aerogel. 

The equilibrium solutions for the quasiclassical
propagator also determine the quasiparticle excitation spectrum below $T_c$.
Continuation of the Matsubara Green's functions to real energies
determines the retarded equilibrium Greens functions,
\be
\hat{g}^R(\hat\vp;\varepsilon) = \hat{\g}(\hat\vp;i\varepsilon_n\rightarrow\varepsilon)
\,,
\ee
which provide spectral information for the low-energy quasiparticle states in the 
presence of pairing correlations and disorder. In particular the angle-resolved 
local density of states, averaged over spin projections, is given by
\be\label{DOS}
N(\hat{\vp};\varepsilon) = -\frac{1}{\pi}\Im g^R(\hat{\vp};\varepsilon)
\,,
\ee
whereas the spin-polarization spectral density is given by 
\be\label{PDOS}
\vP(\hat{\vp};\varepsilon) = -\frac{1}{\pi}\Im \vg^R(\hat{\vp};\varepsilon)
\,.
\ee

\bigskip
\begin{figure}[h]
\begin{center}
\begin{minipage}{0.85\hsize}
\centerline{\psfig{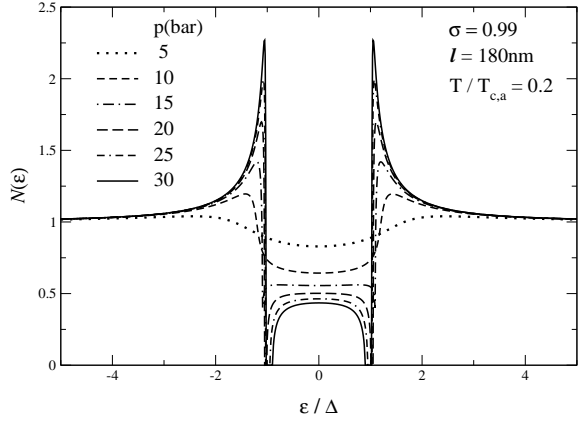}}
\caption{\small The density of states for the BW state in homogeneous aerogel 
         with a mean-free path of $\ell = 1800\,\AA$ and cross-section of 
         $\bar{\sigma}=0.99$. The DOS spectra correspond to pressures
         of  $P=5,10,15,20,25,30\,\mbox{bar}$, and were calculated at the
         same reduced temperature for each pressure, $T=0.2T_{c}$.
	 }
\label{DOS-BW}
\end{minipage}
\end{center}
\end{figure}

Figure \ref{DOS-BW} shows the self-consistently determined density of states (DOS)
for $^3$He-B-aerogel calculated for $\ell = 1800\,\AA$ and $\bar{\sigma}=0.99$ as
a function of energy for pressures spanning the range $P=5-30\,\mbox{bar}$. The 
pairbreaking parameter
\be
\alpha = \frac{\hbar}{2\pi\tau T_{c0}} = \frac{\xi_0}{\ell}
\,,
\ee
decreases from $\alpha = 0.23\rightarrow 0.09$ over the same pressure range. Note
that even at the highest pressure, where the pairbreaking effect is weakest, there
is a substantial density of quasiparticle states at the Fermi level from the 
impurity-induced Andreev bound states. At lower pressures the B-phase becomes 
completely gapless. These quasiparticle states lead to an increase 
in the spin susceptibility.  Their polarizability is described by
the polarization spectral density. In particular, the change in magnetization
can be expressed in terms of $\vP(\hat\vp;\varepsilon)$ by 
analytic continuation of Eq.(\ref{depolarization}) to real energies,
\be
\vm = 2 \int_{-\infty}^{+\infty}\,d\varepsilon\,
f(\varepsilon)\,\langle\vP(\hat\vp;\varepsilon)\rangle
\,,
\ee
where $f(\varepsilon)$ is the Fermi function.

\subsection{Linear Response to the Nuclear Zeeman coupling}

The linear response of $^3$He-aerogel to the nuclear Zeeman field is obtained
by linearizing the propagator in the external field,
\be
\label{linear-expansion}
\hat\g = \hat\g_0 + \hat\g_1 + ...
\,,
\ee
and similarly for the order parameter and self-energies; $\hat\g_0$ is the zero-field 
propagator given in Eq. (\ref{g0_solution}). The first-order correction, $\hat\g_1$, 
satisfies the linearized transport equation,
\be
\label{QCtransport_linear}
\left[i\varepsilon_n\hat\tau_3-\hat\Sig_{\mbox{\tiny imp,0}}-\hat\Delta_0\,,\,\hat{\g}_1\right]
=
\left[\hat{v}_{\mbox{\tiny ext}}+ \hat\Sig_{\mbox{\tiny 1}}+\hat\Delta_1\,,\,\hat{\g}_0\right]
\,,
\ee
and the corresponding linearized normalization conditions,
\be\label{normalization_linear}
\hat\g_0\hat\g_1 + \hat\g_1\hat\g_0 =0\,,\qquad\mbox{with}\quad \hat\g_0^2=-\pi^2\hat{1}
\,.
\ee
The operators on the left-side of Eq. (\ref{QCtransport_linear})
can be expressed in terms of $\hat\g_0$,
\be
i\varepsilon_n\hat\tau_3-\hat\Sig_{\mbox{\tiny imp,0}}-\hat\Delta_0
=
N_0^{-1}\hat\g_0 - n_{\mbox{\tiny imp}} t_1\,\hat{1}
\,,
\ee
with $N_0=-\pi/\sqrt{\tilde\varepsilon_n^2+\Delta_0^2}$. Thus, Eq. (\ref{QCtransport_linear}) becomes,
\be\label{QCtransport_linear2}
\left[\hat{\g}_0\,,\,\hat{\g}_1\right]
=
N_0\left[\hat{v}_{\mbox{\tiny ext}}+ \hat\Sig_{\mbox{\tiny 1}}+\hat\Delta_1\,,\,\hat{\g}_0\right]
\,.
\ee
Combining Eq. (\ref{QCtransport_linear2}) with the normalization conditions 
in Eq. (\ref{normalization_linear}) gives the linear correction to the matrix propagator 
in terms of the external field, first-order self-energies and zero-field propagator,
\be\label{g1_matrix}
\hat\g_1 = \frac{1}{2\pi\sqrt{\tilde\varepsilon_n^2+\Delta_0^2}}\,
            \left\{\pi^2\left(\hat{v}_{\mbox{\tiny ext}}+ \hat\Sig_{\mbox{\tiny 1}}+\hat\Delta_1\right)
          + \hat\g_0\left(\hat{v}_{\mbox{\tiny ext}}+ \hat\Sig_{\mbox{\tiny 1}}+\hat\Delta_1\right)\hat\g_0
             \right\}
\,.
\ee
The detailed solution for $\hat\g_1$ is most conveniently represented in terms of components of $\hat\g_1$
which are symmetrized between the particle and hole components, i.e. 
for any particular component of $\hat\g_1$ define $X^{\pm}\equiv X \pm \underline{X}$.
The symmetry relations (Eqs. \ref{gsymmetry}-\ref{fsymmetry}) imply that $X^{\pm}$ will have 
even or odd symmetry with respect to 
$(\hat\vp,\varepsilon_n)\rightarrow(-\hat\vp,-\varepsilon_n)$
or
$(\hat\vp,\varepsilon_n)\rightarrow(-\hat\vp,\varepsilon_n)$.
 
The self-consistent solution for $\hat\g_1$ requires the first-order corrections
for the internal field, the order parameter, and self-energies.
We can express the correction to the impurity self-energy
in terms of the first-order correction to the t-matrix,
\be
\hat{t}_1=N_f\hat{t}_0\langle\hat\g_1\rangle\hat{t}_0
\,,
\ee
where $\langle\dots\rangle$ denotes the average over the Fermi surface.
Equations (\ref{t0}-\ref{t03}) and (\ref{g1_matrix}) for $\hat{t}_0$ and 
$\hat\g_1$ determine the first-order corrections to the
impurity self-energy, which have the general structure given in Eq. (\ref{Sigma_imp}).
In terms of the symmetrized components, the linear corrections to the impurity
self energy have the form,
\ber
\label{Sigma_imp1}
\Sig^{+}_{\mbox{\tiny imp,1}} &=& n_{\mbox{\tiny imp}}\,N_f\,\left\{\left(t_1^2 + t_3^2\right)\langle g^{+}_1\rangle
                                              + 2t_1 t_3\,\langle g^{-}_1\rangle\right\}
\,,
\\
\Sig^{-}_{\mbox{\tiny imp,1}} &=& n_{\mbox{\tiny imp}}\,N_f\,\left\{\left(t_1^2 + t_3^2\right)\langle g^{-}_1\rangle
                                              + 2t_1 t_3\,\langle g^{+}_1\rangle\right\}
\,,
\\
\label{h_imp1+}
\vh^{+}_{\mbox{\tiny imp,1}} &=& n_{\mbox{\tiny imp}}\,N_f\,\left\{\left(t_1^2 + t_3^2\right)\langle \vg^{+}_1\rangle
                                              + 2t_1 t_3\,\langle \vg^{-}_1\rangle\right\}
\,,
\\
\label{h_imp1-}
\vh^{-}_{\mbox{\tiny imp,1}} &=& n_{\mbox{\tiny imp}}\,N_f\,\left\{\left(t_1^2 + t_3^2\right)\langle \vg^{-}_1\rangle
                                              - 2t_1 t_3\,\langle \vg^{+}_1\rangle\right\}
\,,
\\
\Delta^{\pm}_{\mbox{\tiny imp,1}} &=& n_{\mbox{\tiny imp}}\,N_f\,\left(t_1^2 - t_3^2\right)\langle f^{\pm}_1\rangle
\,,
\\
\label{vDelta_imp1}
\vDelta^{\pm}_{\mbox{\tiny imp,1}} &=& n_{\mbox{\tiny imp}}\,N_f\,\left(t_1^2 - t_3^2\right)\langle \vf^{\pm}_1\rangle
\,.
\eer
The key corrections are the internal fields, $\vh_{\mbox{\tiny imp,1}}^{\pm}$, and the impurity
correction to the order parameter, $\vDelta^{\pm}_{\mbox{\tiny imp,1}}$. The scalar corrections,
$\Sig^{\pm}_{\mbox{\tiny imp,1}}$, turn out to vanish.

The external field also induces an internal exchange field given 
by Eq. (\ref{FL-vector}). Thus, the total effective field is 
\ber
\tilde{\vh}^+(\hat\vp;\varepsilon_n) &=& \vh^+_{\mbox{\tiny imp,1}}(\varepsilon_n)
                             + \vh^+_{\mbox{\tiny FL,1}}(\hat\vp)
                             + 2\vh_{\mbox{\tiny ext}} 
\,,
\\
\tilde{\vh}^-(\hat\vp;\varepsilon_n) &=&  \vh^-_{\mbox{\tiny imp,1}}(\varepsilon_n)
                             + \vh^-_{\mbox{\tiny FL,1}}(\hat\vp)
\,,
\eer
with the exchange terms given by
\be\label{FL-vector1}
\vh^\pm_{\mbox{\tiny FL,1}}(\hat\vp) = \int\frac{d\Omega_{\hat\vp'}}{4\pi}\,
                                       A^a(\hat\vp\cdot\hat\vp')\,
					T\sum_{\varepsilon_n}\,\vg^\pm_{1}(\hat\vp';\varepsilon_n)
\,.
\ee

In general there is also a linear correction to the spin-triplet order parameter,
\be
\vDelta_1^{\pm}(\hat\vp) = \int\frac{d\Omega_{\hat\vp'}}{4\pi}\,V^t(\hat\vp\cdot\hat\vp')\,
                           T\sum_{\varepsilon_n}\,\vf^{\pm}_1(\hat\vp',\varepsilon_n)
\,,
\ee
which may be combined with the off-diagonal impurity correction,
$\tilde{\vDelta}_1^{\pm}(\hat\vp;\varepsilon_n)=
\vDelta_1^{\pm}(\hat\vp)+\vDelta_{\mbox{\tiny imp,1}}^{\pm}(\varepsilon_n)$.
The full solution for $\hat\g_1$ can be expressed in terms of these effective fields:
\ber
\label{g1+}
g_1^+ &=& 0 \,,
\\
\label{g1-}
g_1^- &=& \frac{\pi}{(\tilde\varepsilon_n^2+\Delta^2)^{3/2}}\,
          i\tilde\varepsilon_n\,\left(\vDelta\cdot\tilde{\vDelta}_1^+\right)\,,
\\
\label{vg1+}
\vec{g_1}^+ &=& \frac{\pi}{(\tilde\varepsilon_n^2 + \Delta^2)^{3/2}}\,
                \vDelta(\vDelta\cdot\tilde{\vh}^+) \,,
\\
\label{vg1-}
\vec{g_1}^- &=& \frac{\pi}{(\tilde\varepsilon_n^2 + \Delta^2)^{3/2}}\,
              \{-\vDelta\times(\vDelta\times\tilde{\vh}^-) + \tilde\varepsilon_n\,\vDelta\times\tilde{\vDelta}_1^- \} \,,
\\
\label{f1+}
f_1^+ &=& \frac{\pi}{(\tilde\varepsilon_n^2 + \Delta^2)^{3/2}}\,(-i\tilde\varepsilon_n)\vDelta\cdot\tilde{\vh}^+ \,,
\\
\label{f1-}
f_1^- &=& 0  \,,
\\
\label{vf1+}
\vec{f_1}^+ &=& \frac{\pi\,\tilde{\vDelta}_1^+}{\sqrt{\tilde\varepsilon_n^2 + \Delta^2}}
               -\frac{\pi}{(\tilde\varepsilon_n^2 + \Delta^2)^{3/2}}\,
                \vDelta\,(\vDelta\cdot\tilde{\vDelta}_1^+) \,,
\\
\label{vf1-}
\vec{f_1}^- &=& \frac{\pi\,\tilde{\vDelta}_1^-}{\sqrt{\tilde\varepsilon_n^2 + \Delta^2}}
              + \frac{\pi}{(\tilde\varepsilon_n^2 + \Delta^2)^{3/2}}\,
              \{-\tilde\varepsilon_n\,\vDelta\times\tilde{\vh}^- 
                + \vDelta\times(\vDelta\times\tilde{\vDelta}_1^-)\} \,.
\eer
We can further simplify the results by noting that $\tilde{\vDelta}_{1}^+$ obeys
homogeneous, linear self-consistency equations which allow only the 
solution $\tilde{\vDelta}_{1}^+ = 0$. Thus, $g_1^-$ and  $\vf_1^+$ also vanish.

The magnetization of $^3$He-B in aerogel can be calculated from the Fermi surface
average of Eq. (\ref{vg1+}) and the corresponding total field. Thus, the susceptibility 
(Eq. (\ref{magnetization})) is given by
\be\label{chiB}
\frac{\chi_B}{\chi_N}= 
1+\frac{1}{2\mu H}\,T\,\sum_{\varepsilon_n}\,\langle \vg_1^+\rangle
\,.
\ee
Note that only the even (+) part of $\vg$ contributes to the magnetization.
We can calculate the change in the magnetization,
\be
\label{m}
\vm = T\,\sum_{\varepsilon_n}\,\langle\vg_1^+\rangle
\,,
\ee
from the solutions for $\vec{g_1}^{\pm}$ in Eqs. (\ref{vg1+}-\ref{vg1-}), and the self-consistency equations
for $\tilde{\vh}^{\pm}$ - Eqs. (\ref{h_imp1+},\ref{h_imp1-},\ref{FL-vector1}) -
and $\tilde{\vDelta}_1^{-}$ given in Eq. (\ref{vDelta_imp1}) and Eq. (\ref{vf1-}).
Inserting the result for $\vg^+_1$ from  Eq. (\ref{vg1+}) gives,
\be\label{m-angleaverage}
\vm=T\,\sum_{\varepsilon_n}\,K^+(\varepsilon_n)\,
       \langle\hat\vp\,(\hat\vp\cdot\tilde{\vh}^+)\rangle
\,,
\ee
where 
\be
K^+(\varepsilon_n)\equiv\frac{\pi\Delta^2}{\left[\tilde{\varepsilon}_n^2 + \Delta^2\right]^{3/2}}
\,.
\ee
The Fermi surface average of the external field and the exchange contributions to the 
total field can be evaluated, or expressed in terms of $\vm$,
\be
\langle\hat\vp\,(\hat\vp\cdot\tilde{\vh}^+)\rangle 
			= \twothirds\vh_{\mbox{\tiny ext}}
			+ \onethird\,A_0^a\,\vm
			+ \twofifteenths\,A_2^a\,\vm 
			+ \onethird\,\vh_{\mbox{\tiny imp}}^+
\,.
\ee
We then obtain an equation for $\vm$ that depends on the impurity contribution to 
the internal field,
\be\label{m-intermediate}
\left(1 -\onethird A_0^a\,y_{\tinythreehalves}
        -\twofifteenths A^a_2\,y_{\tinythreehalves}\right)\vm
=
\twothirds\,y_{\tinythreehalves}\,\vh_{\mbox{\tiny ext}}
+
\onethird\,T\,\sum_{\varepsilon_n}\,K^+(\varepsilon_n)\,\vh_{\mbox{\tiny imp}}^+
\,,
\ee
where
\be
y_{\tinythreehalves} = T\,\sum_{\varepsilon_n}\,K^+(\varepsilon_n) 
\ee
is related to the Yosida function ($Y=1-y_{\tinythreehalves}$) 
evaluated with renormalized Matsubara frequencies.
To complete the calculation we need to calculate
the sum over the impurity correction to the internal field, 
$\vh^+_{\mbox{\tiny imp}}$. In general, for scattering 
that is neither in the Born or unitarity limits, the self-consistent solution
for the impurity field entering Eq. (\ref{m-intermediate}) involves both
the even- and odd components of the quasiparticle polarization, $\vg_1^{\pm}$
(see Eq. (\ref{h_imp1+})). However, for the Born and unitarity 
scattering limits the coupling to $\vg_1^{-}$
drops out, and an analytic result can be obtained for 
$\vh^+_{\mbox{\tiny imp}}(\varepsilon_n)$.

In the unitarity scattering limit, $\bar\sigma\rightarrow 1$, 
in which case $t_1 \rightarrow 0$ and 
\be
\label{unitarity-t3}
t_3\rightarrow \frac{1}{\pi N_f}\,\frac{\sqrt{\tilde\varepsilon_n^2 + \Delta^2}}{i\tilde\varepsilon_n}
\,.
\ee
Thus, the equation for $\vh^+_{\mbox{\tiny imp}}(\varepsilon_n)$:
\be\label{h_impurity}
\vh_{\mbox{\tiny imp}}^+ = -\frac{1}{2\pi\tau}
               \left(\frac{\tilde{\varepsilon}_n^2 + \Delta^2}{\tilde{\varepsilon}_n^2}\right)\,
               K^{+}(\varepsilon_n)\,
               \langle\hat\vp\,(\hat\vp\cdot\tilde{\vh}^+)\rangle 
\,,
\ee
depends on the same Fermi surface average that determines $\vm$ in Eq. (\ref{m-angleaverage}), in which
case we can solve for $\vh^+_{\mbox{\tiny imp}}$, and then 
$T\sum_{\varepsilon_n}\,K^+(\varepsilon_n)\,\vh^+_{\mbox{\tiny imp}}(\varepsilon_n)$,
in terms of $\vh_{\mbox{\tiny ext}}$ and $\vm$:
\be
\vh^+_{\mbox{\tiny imp}} = -\frac{1}{2\pi\tau}
\frac{K^+}{\left(\frac{\tilde{\varepsilon}_n^2}{\tilde{\varepsilon}_n^2 + \Delta^2}\right) + \frac{1}{6\pi\tau}\,K^+}
\,\left[\twothirds\vh_{\mbox{\tiny ext}} + \left(\onethird\,A_0^a +\twofifteenths\,A_2^a\right)\vm \right]
\,.
\ee
The result for the change in magnetization reduces to 
\be
\label{m-final}
\vm = \frac{\twothirds\,\tilde{y}_{\tinythreehalves}\,\vh_{\mbox{\tiny ext}}}
     {\left[1-\onethird A^a_0\,\tilde{y}_{\tinythreehalves}-\twofifteenths A_2^a\,\tilde{y}_{\tinythreehalves}\right]}
\,,
\ee
where the impurity-renormalized response function is given by
\be
\label{ytilda}
\tilde{y}_{\tinythreehalves}= \pi T\,\sum_{\varepsilon_n}\,
       \frac{\Delta^2}{[\tilde\varepsilon_n^2 + \Delta^2]^{3/2}}\,
       \left\{\frac{1}{1+\frac{1}{6\tau}\left(\frac{\Delta^2}{\tilde\varepsilon_n^2}\right)
                                              \frac{1}{\sqrt{\tilde\varepsilon_n^2 + \Delta^2}}}\right\}
\quad\mbox{(unitarity)}
\,.
\ee
Thus, the susceptibility of the BW phase of $^3$He in aerogel 
in the HSM is obtained from Eqs. (\ref{chiB}), (\ref{m}) and (\ref{m-final}),
\be
\label{chiB-penultimate}
\frac{\chi_B}{\chi_N} = 1 - \left(\frac{1}{1+F_0^a}\right)\,
   \left(\frac{\onethird\tilde{y}_{\tinythreehalves}}
  {1-\left(\onethird A_0^a + \twofifteenths A_2^a\right)\tilde{y}_{\tinythreehalves}}\right)
\,.
\ee
This result can be expressed in terms of the conventional Landau parameters, $F_0^a$ and $F_2^a$,
\be
\label{chiB-final}
\chi_B/\chi_N = \frac{(1+F_0^a)\left[\twothirds + \tilde{Y}(\onethird + \onefifth F^a_2)\right]}
                     {1+F_0^a(\twothirds+\onethird \tilde{Y})
		     +\onefifth F^a_2(\onethird+\twothirds \tilde{Y})
                     +\onefifth F^a_2 F^a_0 \tilde{Y}}
\,,
\ee
where $\tilde{Y}=1-\tilde{y}_{\tinythreehalves}$ is the impurity renormalized Yosida function
given in Eq. (\ref{Y-unitarity}).

\subsubsection*{Born limit}

In the Born limit, $\bar\sigma \rightarrow 0$, we can carry through essentially the 
same analysis as in the unitarity case to calculate the impurity contribution to the 
local field and obtain,
\be
\vh^+_{\mbox{\tiny imp}} = \frac{1}{2\pi\tau}
\left(\frac{K^+}{1 - \frac{1}{6\pi\tau}\,K^+}\right)
\,\left[\twothirds\vh_{\mbox{\tiny ext}} + \left(\onethird\,A_0^a +\twofifteenths\,A_2^a\right)\vm \right]
\,.
\ee
For the Born limit the susceptibility is again given by (\ref{chiB-penultimate}), but with 
\be
\label{ytilde-Born}
\tilde{y}_{\tinythreehalves}= \pi T\,\sum_{\varepsilon_n}\,
       \frac{\Delta^2}{\left[\tilde\varepsilon_n^2 + \Delta^2\right]^{3/2}}\,
       \left\{\frac{1}{1-\frac{1}{6\tau}\frac{\Delta^2}{\left[\tilde\varepsilon_n^2+\Delta^2\right]^{3/2}}}\right\}
\quad\mbox{(Born)}
\,,
\ee
which yields Eq. (\ref{Y-Born}).

\subsection{Ginzburg-Landau Limit}\label{GLResults}

For temperatures just below $T_c$ the mean-field spin susceptibility
for non-ESP states decreases linearly with temperature. The 
temperature interval over which the susceptibility is linear 
defines the region of applicability of the Ginzburg-Landau (GL) theory. The general
result for the susceptibility of the B phase in the GL region is expressed 
in terms of the GL material parameter for the change in the Zeeman energy,
i.e. $\Delta\Omega_{\mbox{\tiny Zeeman}} = g_z\,\Delta^2\, H^2$,\cite{vol90} 
which implies
\be
\label{chiB-GL}
\chi_B - \chi_N = -2\,g_z \Delta(T)^2 \sim - g_z\,T_c\,(T_c - T)
\,.
\ee
We can obtain the GL limit from the general result for $\chi_B$ and identify
the GL material parameter $g_z$ for the HSM of $^3$He-aerogel.
For $T\rightarrow T_c$, the spin susceptibility reduces to
\be
\frac{\chi_B}{\chi_N} = 1 - \onethird\left(\frac{1}{1+F_0^a}\right)\,\tilde{y}_{\tinythreehalves}
\,,
\ee
with 
\be
\tilde{y}_{\tinythreehalves}\rightarrow\left(\frac{\Delta(T)}{2\pi T_c}\right)^2
\sum_{n\ge 0}\left(\frac{1}{n + \tinyonehalf + \tinyonehalf \frac{1}{2\pi\tau T_c}}\right)^3
\,.
\ee
Thus, the spin susceptibility reduces to
\be
\frac{\chi_B}{\chi_N} = 1 - \onethird\left(\frac{1}{1+F_0^a}\right)\,\left(\frac{\Delta(T)}{2\pi T_c}\right)^2
\,S_3(x)
\,,
\ee
where
\be\label{Sums}
S_m(x) \equiv \sum_{n\ge 0} \left(\frac{1}{n + \tinyonehalf + x}\right)^m \,,\quad m > 1
\,,
\ee
and $x=1/4\pi\tau T_c$ is the pairbreaking parameter.
The order parameter in the GL limit can also be expressed in terms of $T_c$
and the sums defined in Eq. (\ref{Sums}),
\be\label{Delta_GL}
\Delta \rightarrow \sqrt{8\pi^2\left(\frac{1 - x S_2(x)}
                                          {S_3(x)+(2\bar{\sigma}-1)\,x S_4(x)}\right)}\,T_c\,\sqrt{1 - T/T_c}
\,.
\ee

The result for the GL material parameter that determines the change in spin susceptibility
is then identified as
\be
g_z = \frac{N_f\mu^2}{(1 + F_0^a)^2} \, \frac{S_3(x)}{12\pi^2 T_c^2}
\,.
\ee
For $x=0$ we recover the known weak-coupling result for $g_z$ for pure $^3$He-B,
\be
g_z^{\mbox{\tiny pure}} = \frac{N_f\mu^2}{(1 + F_0^a)^2} \, \frac{7\zeta(3)}{12\pi^2 T_{c0}^2}
\,.
\ee

\section{Susceptibility}\label{Results}

\begin{figure}[h]
\begin{center}
\begin{minipage}{0.85\hsize}
\centerline{\psfig{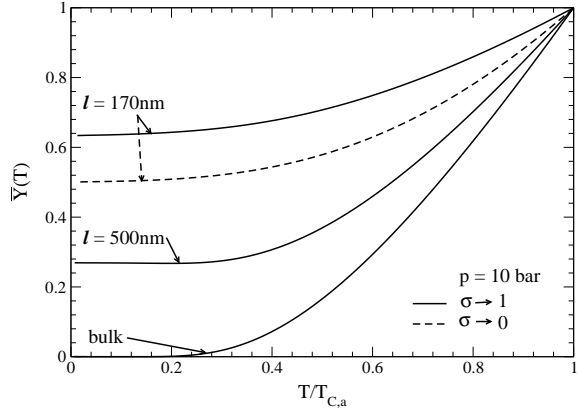}}
\caption{\small The renormalized Yosida function for the spin response 
         function for the BW state vs. $T/T_{ca}$ for
         the unitarity limit (solid lines) of the HSM for $P=10\,\mbox{bar}$ 
	 and $\ell=1700\,\AA$, $\ell=5000\,\AA$ and $\ell=\infty$ (bulk).
         The result for the Born limit (dashed line) is shown for $\ell=1700\,\AA$.
	 }
\label{Y-aerogel_10}
\end{minipage}
\end{center}
\end{figure}

In Figure \ref{Y-aerogel_10} we show the impurity-renormalized 
Yosida function, $\tilde{Y}$ at $P=10\,\mbox{bar}$ in the unitarity limit of the HSM
for mean-free paths of $\ell=1700\,\AA$, $\ell=5000\,\AA$ and $\ell\rightarrow\infty$.
The increase in $\tilde{Y}(T=0)$ reflects the density of fermionic states 
below the gap that can contribute to the low-field magnetization. 
We also show the result for $\tilde{Y}$ in the Born scattering limit at $\ell=1700\,\AA$. 
The smaller cross-section is a less effective pairbreaker, which is reflected in the
stronger condensate suppression of the spin-polarization response function.

\vspace{5mm}

\begin{figure}[h!]
\begin{center}
\begin{minipage}{0.85\hsize}
\centerline{\psfig{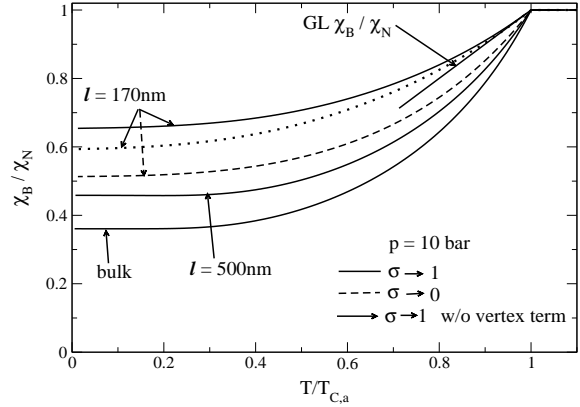}}
\caption{\small The susceptibility for the BW state vs. $T/T_{ca}$ scaled to $\chi_N$ in 
         the unitarity limit (solid lines) of the HSM for $P=10\,\mbox{bar}$ 
	 and $\ell=1700\,\AA$, $\ell=5000\,\AA$ and $\ell=\infty$ (bulk). 
	 The GL slope for $\ell =1700\,\AA$ is plotted to indicate 
         the limited range of validity of the GL result. Also shown for $\ell=1700\,\AA$
         is the effect of ``switching off'' the impurity correction to the local field
         in the unitarity limit (dotted curve).
         The Born limit for $\chi_B$ is shown as the dashed curve for $\ell=1700\,\AA$.
	 The material parameters used for $P=10\,\mbox{bar}$ are: $v_f = 45.66\,\mbox{m/s}$, 
	 $T_{c0}= 1.83\,\mbox{mK}$, $F_0^a=-0.744$, $F_2^a=-0.466$.
	 }
\label{chi-aerogel}
\end{minipage}
\end{center}
\end{figure}

The magnetic susceptibility for $^3$He-B in aerogel is shown in Fig. \ref{chi-aerogel}
for the same pressure, scattering cross-sections and mean-free paths.
The Ginzburg-Landau limit for unitarity scattering and $\ell=1700\,\AA$ is also shown.
The Fermi-liquid parameters are taken from the tables in the review article by 
Halperin and Varoquaux.\cite{hal90} The effect of scattering by the aerogel is substantial;
$T_c$ is suppressed to $T_c \approx 0.5 T_{c0}$ at $P=10,\mbox{bar}$ while the susceptibility
is increased by nearly a factor of $2$ at low temperatures. Another feature to note is
that the Ginzburg-Landau limit for $\chi_B$ is valid only in a limited range of temperatures
below $T_c$; thus, linear fits to limited data for $\chi(T)$ may lead to misleading information
on the material parameters when compared with the GL result.

\vspace*{1cm}
\begin{figure}[h!]
\begin{center}
\begin{minipage}{0.85\hsize}
\centerline{\psfig{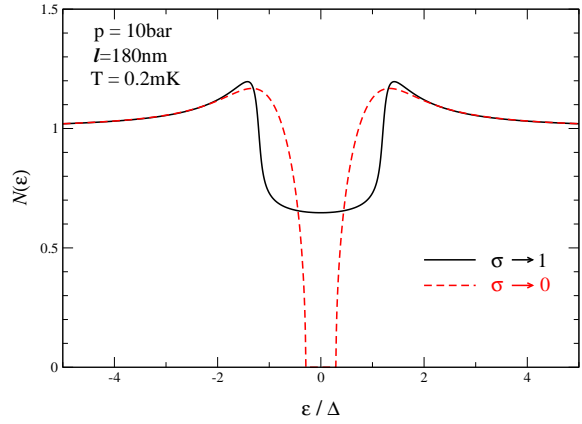}}
\caption{\small The density of states for the BW state vs. 
         energy, $\varepsilon$, for both the unitarity and Born
         scattering limits in the HSM for $P=10\,\mbox{bar}$ 
	 and $\ell=1800\,\AA$.
	 }
\label{DOS_spectrum}
\end{minipage}
\end{center}
\end{figure}

\begin{figure}[h!]
\begin{center}
\begin{minipage}{0.85\hsize}
\centerline{\psfig{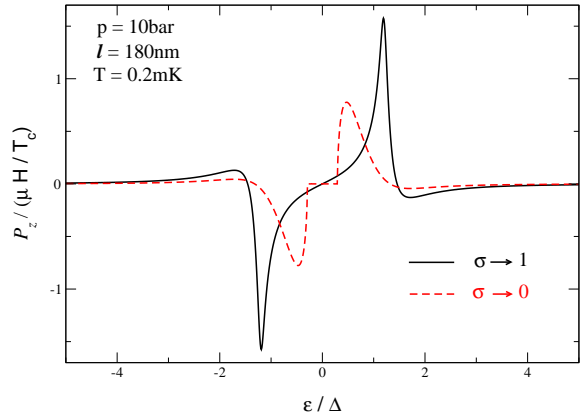}}
\caption{\small The spin polarization spectral function for the BW state vs. 
         energy, $\varepsilon$, for both the unitarity and Born
         scattering limits in the HSM for $P=10\,\mbox{bar}$ 
	 and $\ell=1800\,\AA$.
	 }
\label{Polarization_spectrum}
\end{minipage}
\end{center}
\end{figure}

\noindent
Also shown in Fig. \ref{chi-aerogel}
for $\ell=1700\,\AA$ is a comparison between the unitarity and Born limits for the low temperature
susceptibility. Unitary scattering is more effective at pairbreaking than Born scattering.
In the same figure we show the effect of the impurity vertex correction to the internal field
(Eq. {\ref{h_impurity})) on the spin susceptibility at $\ell=1700\,\AA$ in the unitarity limit;
the impurity field leads to a relatively small correction compared to the overall increase 
in the susceptibility that results from the polarizability of the quasiparticle states.
The large density of states near zero energy (see Fig. \ref{DOS_spectrum}),
characteristic of the unitarity limit, contributes relatively little to the polarization, 
as is evident from Fig. \ref{Polarization_spectrum}. As a result the quasiparticles 
states at finite energy below and near the gap edge, which are present for both unitarity 
and Born scattering are mainly responsible for the increase in the spin susceptibility. 
The integrated polarization from the gapless states below the Fermi level is larger 
in the unitarity limit reflecting the stronger pairbreaking effect
for scattering from larger cross-section impurities.

\subsection{Comparison with Existing Data for the Susceptibility}

Experimental data for the nuclear spin susceptibility of $^3$He in 
aerogel is reported by Sprague, et al.\cite{spr95} for $P=18.7\,\mbox{bar}$
and by Barker, et al.\cite{bar00} for $P=32\,\mbox{bar}$. In both
experiments the susceptibility was reported to decrease below $T_{ca}$
when several monolayers of $^4$He were added to suppress the solid-layer Curie
susceptibility.

\vspace*{1cm}
\begin{figure}[h]
\begin{center}
\begin{minipage}{0.85\hsize}
\centerline{\psfig{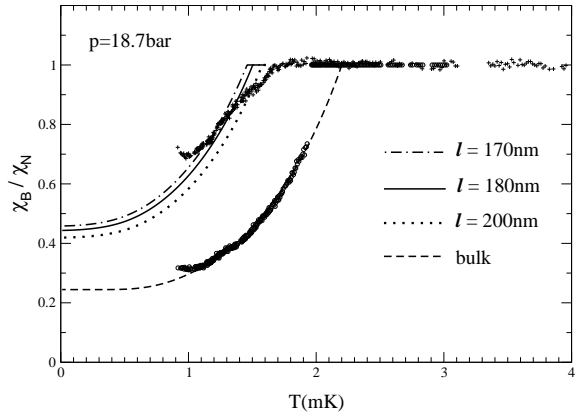}}
\caption{\small Magnetic susceptibility of $^3$He in aerogel at $P=18.7\,\mbox{bar}$
                from Sprague, et al.  The bulk susceptibility data (open circles) is 
		also shown for the same pressure. The Landau parameter $F_0^a=-0.757$ 
		is taken from the normal state susceptibility, while $F_2^a$ is used to fit the 
		bulk data to the theoretical result. The fit parameter is: $F_2^a = 2.2$.
		The theoretical results for $^3$He-aerogel are shown for the same
		pressure and Landau parameters for 
		mean free paths of $\ell = 1700\,,1800\,, \mbox{and}\, 2000\,\AA$, 
		and $\bar{\sigma}=1$.
	 }
\label{NU_Data}
\end{minipage}
\end{center}
\end{figure}

In Fig. \ref{NU_Data} we compare the susceptibility data at $P=18.7\,\mbox{bar}$
with the theoretical result calculated for the same pressure in the unitarity
scattering limit. The comparison is only partly satisfactory. 
The pure $^3$He-B susceptibility reported for the
same pressure by Sprague, et al.\cite{spr95} does not agree precisely with the 
theoretical result using the Fermi-liquid parameters based on the susceptibility
measurements of Hoyt, et al.\cite{hoy81} However, if we fix $F_0^a$ by the normal
state susceptibility, use the Landau parameter $F_2^a$ 
to fit the bulk susceptibility data shown in Fig. \ref{NU_Data}, calculate 
the susceptibility for $^3$He-aerogel with these Landau parameters,
then the comparison with the data for $^3$He-aerogel is qualitatively correct for 
reasonable values of the mean free path.

The comparison between the HSM for $^3$He-B in aerogel and the data of
Barker et al. is slightly improved. The theoretical calculation shown in 
Fig. \ref{Stanford_Data} was obtained with the bulk Fermi-liquid data from
Halperin and Varoquaux's tables, and a mean-free path of $\ell = 1800\,\AA$
for unitarity scattering. The data of Barker, et al. agrees semi-quantitatively
with the theoretical result, but the data span a limited temperature range 
and shows an unphysical upturn at lower temperatures, which is also
visible in the data of Sprague, et al. even for the bulk measurements.

\vspace*{1cm}

\begin{figure}[h!]
\begin{center}
\begin{minipage}{0.85\hsize}
\centerline{\psfig{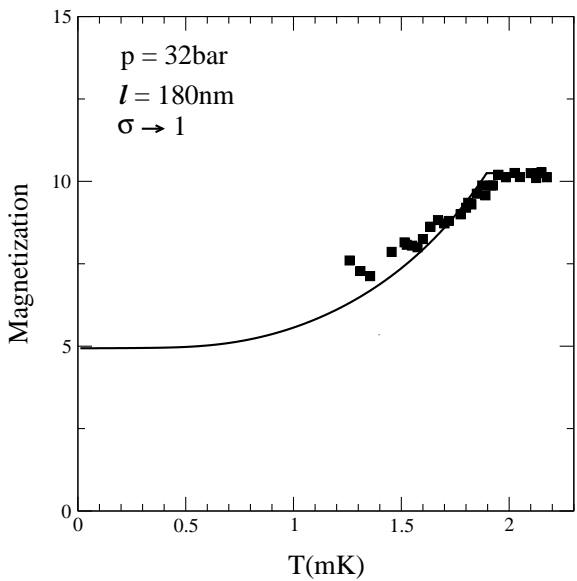}}
\caption{\small Magnetic susceptibility of $^3$He in aerogel at $P=32\,\mbox{bar}$
                from Barker, et al. The theoretical result calculated for the same
		pressure, a mean free path of $\ell = 1800\,\AA$ for $\bar{\sigma}=1$.
	 }
\label{Stanford_Data}
\end{minipage}
\end{center}
\end{figure}

\section{Conclusion}

Experiments based on NMR and acoustic attenuation indicate that the
equilibrium phase of superfluid $^3$He-aerogel at all pressures
is a non-ESP phase, generally assumed to be the Balian-Werthamer
state modified by scattering off the aerogel structure. For $^3$He-aerogel
impurity scattering leads to substantial changes in the susceptibility 
of the BW phase. The available data is in semi-quantitative agreement with
theoretical predictions. Additional measurements could prove
important for a more definitive identification of the equilibrium
order parameter, as well as for refining the theoretical model
to describe the effects of disorder and scattering on the properties of
superfluid $^3$He.

\section{Acknowledgement}

We thank Bill Halperin, Tom Haard, and Guillaume Gervais
for enlightening discussions on the phase diagram and properties of
superfluid $^3$He in aerogel. This research was supported in part
by the NSF through grant DMR-9972087.


\end{document}